\documentclass[graybox]{svmult}

\usepackage{mathptmx}       
\usepackage{helvet}         
\usepackage{courier}        
\usepackage{type1cm}        
\usepackage{makeidx}         
\usepackage{graphicx}        
\usepackage{multicol}        
\usepackage[bottom]{footmisc}
\usepackage{amssymb}
\usepackage{amsmath}

\newcommand{\ba}{\begin{eqnarray}}
\newcommand{\ea}{\end{eqnarray}}
\newcommand{\be}{\begin{equation}}
\newcommand{\ee}{\end{equation}}
\newcommand{\bi}{\begin{itemize}}
\newcommand{\ei}{\end{itemize}}

\newcommand{\da}{\delta}
\newcommand{\la}{\lambda}
\newcommand{\ka}{\kappa}

\newcommand{\sa}{\sigma}

\newcommand{\La}{\Lambda}

\newcommand{\cF}{{\cal F}}
\newcommand{\cP}{{\cal P}}
\newcommand{\cQ}{{\cal Q}}
\newcommand{\cR}{{\cal R}}
\newcommand{\cO}{{\cal O}}

\newcommand{\cS}{{\cal S}}
\newcommand{\cT}{{\cal T}}

\newcommand{\w}{\widetilde}

\newcommand{\p}{\partial}
\newcommand{\hp}{h^{\perp}}
\newcommand{\Ap}{A^{\perp}}
\newcommand{\n}{\nabla}

\newcommand{\ra}{\rightarrow}

\newcommand{\LF}{\left(}
\newcommand{\RF}{\right)}
\newcommand{\LT}{\left[}
\newcommand{\RT}{\right]}
\newcommand{\Ld}{\left.}
\newcommand{\Rd}{\right.}

\newcommand{\hw}{\w{h}}

\newcommand{\gb}{\bar{g}}
\newcommand{\Rb}{\bar{R}}

\newcommand{\2}{\frac{1}{2}}

\newcommand{\4}{\frac{1}{4}}

\newcommand{\mx}{\mbox}
\newcommand{\mt}{\mathtt}
\newcommand{\mand}{\mx{ and }}

\newcommand{\ie}{{\it i.e.,\ }}

\newcommand{\non}{\nonumber\\}

\newcommand{\Fc}{\mathcal{F}}

\newcommand{\hco}{\hat{\mathcal{O}}}

\makeindex             

\begin{document}

\title*{Gravitational theories with stable (anti-)de Sitter backgrounds}
\author{Tirthabir Biswas, Alexey S. Koshelev and Anupam Mazumdar}
\institute{Tirthabir Biswas \at Loyola University, 6363 St. Charles Avenue, Box 92, New Orleans 70118, USA\\ \email{tbiswas@loyno.edu}
\and Alexey S. Koshelev \at Departamento  de F\'isica and Centro  de  Matem\'atica  e
Aplica\c c\~oes,  Universidade  da  Beira  Interior,  6200  Covilh\~a,
Portugal\\ \email{alexey@ubi.pt}
\and Anupam Mazumdar \at Consortium for Fundamental Physics, Lancaster University, LA1 4YB, UK\\ \email{a.mazumdar@lancaster.ac.uk}}
%
%
\maketitle

\abstract{In this article we will construct the most general torsion-free parity-invariant covariant theory of gravity that is
free from ghost-like and tachyonic instabilities  around  constant curvature space-times in four dimensions. Specifically, this includes the Minkowski, de Sitter and anti-de Sitter backgrounds. We will first argue in details how starting from a general covariant action for the metric one arrives at an ``equivalent'' action that at most contains  terms that are quadratic in curvatures but nevertheless is sufficient for the purpose of studying stability of the original action. We will then briefly discuss how such a ``quadratic curvature action'' can be decomposed in a covariant formalism into separate sectors involving the  tensor, vector and scalar modes of the metric tensor; most of the  details of the analysis however, will be presented in an accompanying paper. We will find that only the
transverse and trace-less spin-2 graviton  with its two helicity states and possibly a spin-0 Brans-Dicke type scalar degree of freedom  are left to propagate in $4$ dimensions. This will also enable us to arrive at the consistency conditions required to make the theory perturbatively stable around constant curvature backgrounds.
\\
This will be included
 as a chapter in the book  entitled ``At the Frontier of Spacetime --
Scalar-Tensor Theory, Bells Inequality, Machs Principle, Exotic
Smoothness''  (Springer 2016)}
\section{Introduction}
\subsection{A personal note from Tirtho}
Initially, it felt a bit strange to me to write about attempts to modify General Theory of Relativity (GR) when we are celebrating one hundred very successful years of Einstein's master piece, but then I remembered one of the fundamental tenants of science, that we can never know whether a theory is correct,  only that it is not yet wrong! So, it is not so surprising after all that in spite of its success  in these hundred years, literally hundreds of attempts have been made to  modify Einstein's theory of gravity. Having said that, GR has proved to be impossibly difficult to dislodge. Perhaps, there is an emotional component to it, after all we all fell in love with GR when we saw for the first time how a theory could replace the abstract notions of force and ``action at a distance" with a  physically intuitive and beautiful geometric picture that could explain gravity. And, it is always hard to extricate ourselves from something that we love. While we now know how to construct countless theories of gravity which preserves the same basic geometric structure and symmetries, for instance, that the force of gravity is encoded in curvatures of space-time that can be built from a metric, Einstein had also based his theory on deep philosophical ideas, such as the Equivalence principle, that are  harder to preserve, not that we are obliged to. Theoretically speaking, if one at least wants to preserve general covariance without introducing any new fields beyond the metric, the modifications that one can consider must  involve higher derivative terms which tend to be plagued by ghost-like or tachyonic instabilities. As if these were not enough impediments, experimentally, GR has been tested to unprecedented levels of accuracy, and it passes with flying colors. Indeed, GR has experienced success in explaining a multitude of different phenomena starting from purely astronomical observations, such as the bending of light near a massive object, to the cosmic expansion of our universe.

So, why do we keep searching for this elusive ``better" theory so vigorously, why can't we just leave GR alone  for a while? The answer is obvious to most  theoretical physicists, notwithstanding its amazing success, GR is profoundly incomplete. It is plagued by classical singularities in the ultraviolet (UV), as seen inside the blackholes or at the Big Bang.  GR suffers from quantum divergences that cannot be renormalized and constructing a consistent quantum theory of gravity remains one of the outstanding challenges of 21st century physics. To draw a contrast, while we may not be completely happy with the Standard Model of particle physics  that describes the three other fundamental forces of nature, for instance, it suffers from the hierarchy problem, doesn't explain the origin of its twenty odd parameters, the fact of the matter is that it is a perfectly consistent theory that has till date explained/predicted experimental observations quite brilliantly.

On the infrared (IR) front we have also ``recently" been greeted with a surprise, we have found out  that our universe at the largest scales is apparently able to defy gravity and speed up its cosmic expansion. While this major inconvenience can be explained away without having to tamper with GR, for instance, just by invoking a cosmological constant, albeit a disconcertingly small one, there is a school of thought that it is gravity that is perhaps becoming weak at cosmic scales thereby allowing our universe to accelerate.

Thus, today it has become especially fashionable to try to modify GR in ways that could address either the UV or IR problems/puzzles, but the reason, in my opinion, why Carl's work with Dicke is phenomenal is not just because they realized the importance of  going beyond GR and constructed in comprehensive detail their scalar tensor model of gravity, but  more  because they did so around a half a century ago! What is also remarkable is that this first attempt to modify GR is arguably still the most fruitful of the modifications that has been considered in the literature. Indeed, Brans-Dicke theories or generalizations thereof are what emerge from fundamental theories such as Kaluza Klein theories, supergravity models and string theory after  compactifications of extra dimensions. It has been phenomenologically the most successful, finding applications in inflation theory and various dark energy models such as quintessence.

Unsurprisingly, I had worked on several different versions of Brans-Dicke theory before I actually met Carl during my job interview at Loyola. Thankfully, I didn't know that I was actually meeting Carl Brans (somehow I missed his profile on the Loyola physics faculty listings) because that would have completely overwhelmed me. It was only halfway through the interview that I realized that I was talking to someone who knew a lot more gravity than I did. Since then, we have become very good friends, his good nature, his humility, and his commitment to rigor is something that I cherish and I am inspired by. So, here is to Carl for showing the path that many others like me could follow. Cheers!\\
\subsection{Towards consistent theories of gravity}
There have been numerous attempts to formulate quantum theory of gravity ~\cite{Veltman:1975vx,dewittQG,DeWitt:2007mi}, such as
string theory (ST)~\cite{Polchinski:1998rr},  Loop Quantum Gravity (LQG)~\cite{Ashtekar}, Causal Set~\cite{Henson:2006kf}, and
asymptotic safety~\cite{Weinberg:1980gg}.  In many of  these approaches gravitational interactions yield non-local operators
where the interactions are spread out over a space-time region. For instance, strings and branes are non-local objects, nonlocality also emerges in string field theory~\cite{Witten:1985cc}, non-commutative geometry~\cite{noncom},  $p$-adic strings~\cite{Freund:1987kt}, zeta strings~\cite{Dragovich:2007wb}, and strings quantized on a random lattice~\cite{Douglas:1989ve,Biswas:2004qu}, for a review, see \cite{Siegel:1988yz}.

A key feature of all these stringy models is the presence of an {\it infinite series of higher-derivative} terms incorporating the non-locality in the form of an {\it exponential kinetic} correction~\cite{Siegel:2003vt,Tseytlin:1995uq,Biswas:2005qr}, or equivalently modifying the graviton propagator
by an {\it entire function}~\cite{Tomboulis,Modesto,BG,Moffat-qg}. Similar infinite-derivative modifications have also been argued to arise in the asymptotic safety approach to quantum gravity~\cite{Krasnov}~\footnote{Finite higher derivative theories suffer from  Ostrogradsky
instabilities, see Ref.~\cite{Eliezer:1989cr}. However, the Ostrogradsky argument relies on having a
highest ``momentum'' associated with the highest derivative in the theory, in which the energy comes as a linear term, as opposed to quadratic.  In a
classical theory this would lead to instability and in  a quantum theory,  this would yield ghosts or extra poles in the
propagator. A classic example is Stelle's $4th$ derivative theory of gravity~\cite{Stelle:1976gce},
which has been argued to be UV finite, but contains massive spin-2 ghost, therefore shows vacuum instabilities.}.

Only very recently, the concrete criteria for any covariant gravitational theory (including infinite-derivative theories) to be free from ghosts and tachyons around the Minkowski vacuum was obtained in Refs.~\cite{Biswas:2011ar,Biswas:2013kla}. The class of action considered were assumed to be free from torsion, have a well defined Newtonian limit and to be parity conserving.  It was also shown in ~\cite{Biswas:2011ar,Biswas:2013kla}, that one could construct theories that have no extra poles in the propagators, so that there are no new degrees of freedom,  ghosts or otherwise. The only dynamically relevant degrees of freedom are the massless gravitons;  the graviton propagator, however, can be modified by a multiplicative {\it entire function}. In particular, one can choose the entire function to correspond to be a gaussian, which would suppress the ultraviolet (UV) modes {\it possibly} making the theory  asymptotically free~\cite{Talaganis:2014ida}, see also~\cite{Tomboulis,Modesto,BG,Moffat-qg} for similar arguments in slightly different models. We should point out that at a classical level it has already been shown that in these infinite derivative gravity (IDG) models one can find  cosmological solutions bereft of singularities~\cite{Biswas:2005qr,Biswas:2010zk,Biswas:2012bp,Biswas:2011ar},
as well as non-singular static, spherically symmetric metrics~\cite{Biswas:2012bp,Frolov:2015,Biswas:2013cha}~\footnote{The action of
Ref.~\cite{Biswas:2011ar} also provides the UV complete Starobinsky inflation~\cite{Biswas:2013dry,Chialva:2014rla,Craps:2014wga}.  Also, it was noted that the gravitational entropy for this action, for a static spherically symmetric background, gets no contribution from the quadratic curvature part~\cite{Conroy:2015wfa}.}.
These classical results corroborates the idea that IDG theories may provide  us with an asymptotically safe/free theory of gravity, for a review on these models, see~\cite{Biswas:2014tua}.

The aim here  is  to go beyond the analysis around the Minkowski vacuum. We would like to find a robust algorithm  to construct
the most general  action of gravity that is ``consistent'' around  constant curvature maximally symmetric space-times, viz. de Sitter (dS) and anti-de Sitter (AdS) and Minkowski. The analysis in~\cite{Biswas:2011ar} essentially gave us constraints that quadratic curvature terms (such as $R^2$, $R\Box R$, etc.) must satisfy must satisfy in order for the theory to be free from instabilities around the Minkowski space-time. However, the higher curvature terms remained completely arbitrary. If however, we believe that the ``ultimate'' theory of gravity must be consistent on any background, then requiring that it be so will provide  constraints on the higher curvature terms. The ultimate hope is that this may provide us with new insights on how to  construct a consistent and finite quantum theory of gravity. Looking at dS/AdS backgrounds is a first step towards this process where we will start with a gravitational action
that is covariant, parity preserving, torsion-free and possesses a well defined Newtonian limit. Our goal will be to study the quadratic
fluctuations around dS and AdS backgrounds. In this article we will argue that for this purpose it is sufficient to study the fluctuations around an ``equivalent'' action which has terms that are at most quadratic in curvatures. This is a crucial simplification which makes it possible to study the dynamics of linearized fluctuations around dS/AdS/Minkowski backgrounds for a very general class of covariant gravity theories.

To briefly outline our analysis, we note that in $4$ space-time dimensions, a priori, there are
a total 10 independent degrees of freedom in the metric, out of which two degrees of freedom are associated with a massless spin-2 field (tensor mode), two more degrees of freedom with a
massless spin-1 field (vector mode), and two spin-0 fields (scalar modes), along with $4$ gauge degrees of freedom. In a companion paper, using a covariant formalism we were able to show that in the equivalent action (and this really means for any action by our previous argument) the dangerous ghost-like vector mode and one of the scalar modes are absent from the theory, as one might expect from Bianchi identities. Further, following  a rather elaborate calculation, in~\cite{companion} we were also able to decompose the equivalent quadratic curvature action into the remaining  propagating degrees of freedom, the spin-2 gravition and  the spin-0 Brans-Dicke scalar~\footnote{Although, Brans and Dicke formulated their theory by adding a new nonminimally coupled scalar field, as is well known, this scalar degree of freedom can be incoropated within the metric degrees of freedom by replacing $R\ra F(R)$ in the gravitaional action~\cite{equivalence}. This is the approach that naturally emerges in our analysis.}
 and obtain the conditions under which the tensor and the scalar mode can be made ghost and tachyon free in dS/AdS.  While we recommend the readers to our companion paper~\cite{companion} for all the details of the derivations leading up to the consistency conditions, in this article we will briefly outline the important results. Indeed, our results will
match  the Minkowski space-time analysis of Ref.~\cite{Biswas:2012bp}, when we let the  cosmological constant vanish.

Let us now begin our discussion by obtaining the most general form of the gravitational action that is relevant for studying the classical and quantum properties of the fluctuations around dS/AdS backgrounds.

\section{Higher Derivative Actions on (anti-)de Sitter Space-times}
\subsection{Obtaining the General form of the Covariant Derivative Structure}

Our aim in this section is to arrive at the most general form of the gravitational action that is relevant for studying classical and quantum properties of the fluctuations around constant curvature backgrounds. While this was already investigated for the Minkowski space-time in $4$ dimensions in~Ref.~\cite{Biswas:2011ar}, here we generalize the analysis to include dS/AdS backgrounds. Now, for investigating theoretical and observational consistencies of gravitational models, often it is sufficient to consider quadratic fluctuations around relevant background metrics, \ie only keep $\cO(h^2)$ terms  in the action, where
$h_{\mu\nu}$ corresponds to fluctuations around the background metric, $\gb_{\mu\nu}$:
\be
g_{\mu\nu}=\gb_{\mu\nu}+h_{\mu\nu}\,.
\ee
In this article we will restrict ourselves to constant curvature,
maximally symmetric space-times, \ie $\gb_{\mu\nu}$ is  dS/AdS or the Minkowski metric.
Keeping this in mind, let us first identify the most general form of a covariant action that we need to consider if we are only interested in keeping the $\cO(h^2)$ terms in the action. Conversely, this will tell us how to obtain the $\cO(h^2)$ action starting from any arbitrary covariant metric theory of gravity. Our arguments will closely resemble what was discussed for Minkowski space-times in~\cite{Biswas:2011ar,Biswas:2013kla} (see~\cite{Conroy:2015wfa} for its generalization to any dimensions), but they will become more intricate for  dS/AdS backgrounds.

As was first noted in~\cite{Biswas:2011ar}, any covariant action with a well defined Minkowski limit can be written as
\be
S=\int d^4x\sqrt{-g}\LT \cP_0+\sum_i \cP_i\prod_I (\hat{\cO}_{iI} \cQ_{iI})\RT
\label{general}
\ee
where $\cP,\cQ$'s are functions of the Riemann and the metric tensor, while the differential operators $\hat{\cO}$'s are made up solely from covariant derivatives, and contains at least one of them. Essentially, any action which admits a Taylor series expansion in covariant derivatives is included in our discussion. However, nonlocal operators such as $\Box^{-1}$ (see for instance~\cite{conroy1overbox}) falls outside the purview of our analysis.

First of all, it is easy to see that even if the $\cQ$'s are complicated functions of the Riemann tensor to begin with, one can always use simple rules of calculus to  break up $\hat{\cO}_{I} \cQ_{I}$ into a sum of terms where each term is of the form $\prod_J (\hat{\cO}_{J} \cR_{J})$, where $\cR_J$'s now represent just the Riemann tensors. We note in passing that if a metric contraction is present inside the $\cQ$'s they can be moved to $\cP$ as the metric is annihilated by the covariant derivatives, $\n_{\mu}g_{\nu\rho}=0$. In other words, without loss of any generality, we can write our action in the form
\be
S=\int d^4x\sqrt{-g}\LT \cP_0+\sum_{i} \cP_{i}\prod_I (\hat{\cO}_{iI} \cR_{iI})\RT\ .
\ee
Purposely, we have not specified the index structure of the differential operators and the curvature tensors. The most useful
property of the maximally symmetric constant curvature space-times is that the Riemann tensor can be
completely expressed in terms of the metric and therefore the covariant derivatives annihilate the Riemann tensor and any functions thereof.
Mathematically,
\be
\hat{\bar{\cO}}\bar{\cR}=\hat{\bar{\cO}}\bar{\cP}=0
\ee
This, in turn, implies that at most we need to consider terms which contain two $\hat{\cO}$'s: If one has a term like $\hco \cR$, then if both $\hco$ and $\cR$ take on the background curvature values term must vanish. This implies that we need to vary at least one of them, and since we are only interested in quadratic variations, at most we can accommodate two such variations. The relevant action then reduces to
\ba
S&=&\int d^4x\sqrt{-g}\LT \cP_0+\sum_{i} \cP_{1i} (\hat{\cO}_{1i} \cR_{1i})(\hat{\cO}_{2i} \cR_{2i})\Rd+\Ld\sum_i \cP_{2i} (\hat{\cO}_{3i} \cR_{3i})\RT
\label{simple1}
\ea
Let us simplify the second term. First consider the situation that $\cP_1$ is just a constant. In this case, applying repeated integration by parts one can convert the term into the form of the last term. So, $\cP_1$ must contain Riemann tensors. In this case, schematically:
$$
\int d^4x\sqrt{-g} \cP_{1i} (\hat{\cO}_{1i} \cR_{1i})(\hat{\cO}_{2i} \cR_{2i})=$$
$$-\int d^4x\sqrt{-g}  \LF{\hat{\cO}_{1i}\over \n} \cR_{1i}\RF \LT(\n\cP_{1i})(\hat{\cO}_{2i} \cR_{2i})+\cP_{1i}(\n\hat{\cO}_{2i} \cR_{2i})\RT .
$$
The first term is a product of three operators (as long as $\hat{\cO}_{i1}$
contains more than one derivatives) and hence do not contribute to the quadratic
fluctuations, and one can  continue to integrate by parts the ``second'' terms to
keep reducing the number of derivatives from $\cO_{i1}$. This process can
continue till we are left with only a single covariant derivative in
$\hat{\cO}_{i1}$. Thus, the relevant action reduces to the form
\ba
S=\int d^4x\sqrt{-g}\LT \cP+\sum [ \cP (\n \cR)(\hat{\cO} \cR)+\cP \hat{\cO} \cR]\RT\,.
\label{simple2}
\ea
We have suppressed the indices, but remind the readers that $\cP$'s are just made up of the metric and Riemann tensors, while $\hat{\cO}$'s are made up of covariant derivatives.

It is convenient to make a last rearrangement.  Since $\cP,\cR$ contains an even number of indices, the $\hco$ appearing in the third term must contain at least two covariant derivatives. Integrating by parts it is then trivial to see that this term can always be recast as the second. Thus our relevant action is of the form
\ba
S=\int d^4x\sqrt{-g}\LT \cP_0+\sum_{i=1}  \cP_i (\n \cR)(\hat{\cO}_i \cR)\RT\,.
\label{semi-final}
\ea
In other words, given any arbitrary higher derivative action which possesses a well defined Minkowski limit, $\cR\ra 0$, we can always obtain an action of the form  ~(\ref{semi-final}) plus additional terms which do not contribute to the quadratic action involving $h_{\mu\nu}$.

\subsection{Constant Curvature Background Solutions}
Before proceeding any further, we need to determine the vacuum solution around which we want to perturb our action. This, in particular will also tell us whether  ~(\ref{semi-final}) provides us
with an dS/AdS or Minkowski solution.  For this question we need to look at
linear variations of the action. However, since all the terms except the first
contain covariant derivatives acting on two curvatures, and covariant derivatives annihilate the background curvatures, linear variations of these terms must
vanish. Thus, we are only left to consider the linear variation of the first
term, \ie $\da (\int d^4x\sqrt{-g}\cP_0(\cR))$. This has already been discussed in previous literature, but for completeness, below we provide a
discussion and the main result.

Firstly, it becomes useful from this point onwards to consider $\cP_0$ as a
function of the scalar curvature, the traceless Ricci tensor (we will refer to this as the TR tensor from here on)
$$
S_{\mu\nu}=R_{\mu\nu}-\4 Rg_{\mu\nu}\,,
$$
and the Weyl tensor
\begin{equation*}
 C^\mu_{\alpha\nu\beta}= R^\mu_{\alpha\nu\beta} -\frac 1{2}(\delta^\mu_\nu
R_{\alpha\beta}-\delta^{\mu}_\beta
R_{\alpha\nu}+R^\mu_\nu{g}_{\alpha\beta}-R^\mu_\beta{g}_{\alpha\nu}
)+\frac R{6} (\delta^\mu_\nu {g}_{\alpha\beta}-\delta^{\mu}_\beta
{g}_{\alpha\nu})\,,
 \end{equation*}
as the latter two are traceless  and vanish on dS/AdS/Minkowski space-times.

The key point is that since the Lagrangian is a scalar quantity, in all the
scalar polynomials that appear in the Lagrangian there cannot be any term that contains
 a single TR or Weyl tensor, there has to be at least two TR tensors, or two Weyl, or one Weyl plus one TR tensor. Otherwise, their indices have to be  contracted with the metric tensor which makes them vanish.  This means that while taking a single variation of any
such scalar polynomial, there will always remain another TR or Weyl tensor
which then has to take on the background value and hence must vanish. To conclude, we only need to worry about the function
\be
\cP_R(R)=\cP_0(R,S= 0,C=0)\ ,
\ee
and the variation of the action (\ref{semi-final}) reduces to
\ba
\da S=\int d^4x \sqrt{-\gb}\LT{h\over 2}\cP_R(R)-{h\over 4}\cP_R'(R) R\RT\ ,
\ea
where we have dropped some total derivatives.
Thus the background curvature, $\Rb$, is determined by the equation
\be
2\cP_R(\Rb)-\Rb \cP_R'(\Rb)=0\ .
\label{backgroundK}
\ee
\subsection{Classification based on Quadratic Curvature Action}
We are now going to perform  a final simplification or rather a classification: For a given action of the form (\ref{semi-final}), we will attempt to find an action which has a much simpler form, but which nevertheless gives the same quadratic (in $h_{\mu\nu}$) action as that of the original action. It will become evident that several different actions of the form (\ref{semi-final}) will have the same simple {\it equivalent} action. Also, if a particular action admits several background curvatures, \ie (\ref{backgroundK}) has more than one solution, then it will have different equivalent actions depending upon the background about which one wants to find the quadratic action.

Having made these clarifications, let us proceed. We first observe that while obtaining the quadratic action for the fluctuations, $\da \sqrt{-g}$ or $\da \cP_i$,  cannot contribute  in the variation of the second term in (\ref{semi-final}), else the covariant derivatives will annihilate the background Riemann tensors. Thus, the quadratic variation must be given by
\be
\da S=\int d^4x\LT\da (\sqrt{-g} \cP_0)+\sum_{i=1} \sqrt{-\gb} \cP_i(\bar{\cR}) \da(\n \cR)\da(\hat{\cO}_i \cR)\RT\,.
\label{variation}
\ee
Now, the background Riemann tensor can be written completely in terms of the
metric, $\cP_i(\bar{\cR})= \w{\cP}_i(\bar{g})$. Then the terms involving
$\cP_i$'s can be simplified as follows:
\ba
\int &d^4x&\sqrt{-\gb} \cP_i(\bar{\cR}) \da(\n \cR)\da(\hat{\cO} \cR)
=\int d^4x\sqrt{-\gb} \da(\n \cR)\da(\w{\cP}_i(\bar{g}_{\mu\nu}) \hat{\cO}
\cR)\, \nonumber \\
\approx\int &d^4x&\sqrt{-\gb} \da(\n \cR)\da(\w{\cP}_i(g_{\mu\nu}) \hat{\cO} \cR)=\int d^4x\sqrt{-\gb} \da(\n \cR)\da(\w{\cO} \cR)
\label{steps}
\ea
What we have shown here is that the quadratic variation of any action of the
form   (\ref{semi-final}) is exactly the same as the variation coming from
an {\it equivalent} action of the form
\be
S=\int d^4x\sqrt{-g}\LT \cP+\sum (\n \cR)(\w{\cO} \cR)\RT\ ,
\label{intermediate}
\ee
where $\w{\cO}$ can be obtained from $\hat{\cO}$ according to
the prescription  above. Therefore, for the purpose of understanding the linearized
fluctuation dynamics, we need only to consider actions of the form (\ref{intermediate}).

Now, these actions were precisely the type of actions that were considered in~\cite{Biswas:2011ar,Biswas:2013kla}, and the
Bianchi identities along with the commutativity of the covariant
derivatives (we are considering a torsionless theory) enable one to recast it
in the following rather simple form:
\be
S=\int d^4x\sqrt{-g}[\cP_0(\cR)+ R {\Fc}_{1}(\Box)R+S_{\mu\nu}
{\Fc}_{2}(\Box)S^{\mu\nu}+ C_{\mu\nu\la\sa}
{\Fc}_{3}(\Box)C^{\mu\nu\la\sa}]\,.
\label{almost}
\ee
where the $\Fc_{i}$'s are of the form
$$
\cF_{i}(\Box)=\sum_{n=1}^{\infty} c_{i,n}\Box^n
$$
We note that although we continue to use the same symbol, $\cP_0$, this term can actually change as one goes over from   (\ref{intermediate}) to (\ref{almost}). More details including illustrative examples will be provided in the companion paper~\cite{companion}.

To complete the reduction, let us focus on the variation of the $\cP_0(R)$ piece,
see~\cite{chiba,solganik} for similar discussions and conclusions. Once more, since both the Weyl and
symmetric tensors vanish on Minkowski/dS/AdS, we can at most have two of those. Moreover,
we also can't have terms containing a single symmetric or Weyl tensor, since
their indices have to necessarily be contracted which makes them vanish, and
by the same token a mixed term with one symmetric and one Weyl also cannot be
non-vanishing. In other words, the only relevant part of $\cP_0$ in action
  (\ref{almost}) that survives is of the form:
\ba
\cP_0=\cP_R(R)+\cP_S(R)S_{\mu\nu}S^{\mu\nu}+
\cP_C(R)C_{\mu\nu\rho\sa}C^{\mu\nu\rho\sa}\ ,
\ea
where
\be
\cP_S(R)=\2\left({\p^2 \cP_0\over \p S_{\mu\nu}\p S^{\mu\nu}}\right)_{S=C=0}\mand \cP_C(R)=\2\left({\p^2 \cP_0\over \p C_{\mu\nu\rho\sa}\p C^{\mu\nu\rho\sa}}\right)_{S=C=0}
\ee
Finally, for the $S$- and $C$-terms the quadratic
variations must originate from $S$ and $C$ tensors, so the $R$ can
take on the background value, $\Rb$. It is also obvious that the $\cP_R(R)$ reduces
around the dS/AdS/Minkowski background to
\begin{equation*}
\cP_R\ra\frac{M_P^2}2R+c_{1,0} R^2-\Lambda\,,
\end{equation*}
where the parameters of the equivalent action are given by
\be
M_P^2=\frac{4}{\Rb}[\cP_R(\Rb)-\2\Rb^2\cP''_R(\Rb)],\quad c_{1,0}=\frac12\cP''_R(\Rb),\quad \Lambda=\cP_R(\Rb)-\2 \Rb^2\cP''_R(\Rb)={M_P^2\Rb\over 4}\,.\label{params}
\ee
The last inequality was indeed expected in  accordance with   (\ref{backgroundK}).

Thus, the equivalent action involving the non-derivative terms are given by
\be
S=\int d^4x\sqrt{-g}\LT\frac{M_P^2}2R+c_{1,0} R^2+c_{2,0}S_{\mu\nu}S^{\mu\nu}+c_{3,0} C_{\mu\nu\la\sa}C^{\mu\nu\la\sa}-\Lambda\RT
\ ,
\ee
where
\be
c_{2,0}=\cP_S(\Rb)\ ,\  c_{3,0}=\cP_C(\Rb)\ ,
\ee
and the other coefficients are given by   (\ref{params}).

To summarise, we have shown that in order to investigate quadratic fluctuations around dS/AdS/Minkowski space-times in a generic gravitational theory, all we need to focus our attention on are actions of the form:
\begin{equation}
S = \int d^4x\ \sqrt{-g}\left[\frac{M_P^2}2 R-\Lambda
+R
\Fc_1(\Box)R+S_{\mu\nu}
\Fc_2(\Box)S^{\mu\nu}+ C_{\mu\nu\la\sa}
\Fc_{3}(\Box)C^{\mu\nu\la\sa}
\right]\,,
\label{properaction}
\end{equation}
where we have now redefined the $\Fc$'s to include the constant terms:
\be
\cF_i(\Box)=\sum_{n=0}^{\infty} c_{i,n}\Box^n\ .
\label{cFs}
\ee
We point out that typically one expects the higher derivative terms to become important at some scale $M\leq M_p$ which can be made explicit by rescaling the  $c_{i,n}$'s and redefining $\cF_i(\Box)\ra\cF_i(\Box/M^2)$. This is especially useful for constructing phenomenological models and will be discussed in~\cite{companion}, here though we will work with (\ref{cFs}).

In the process of arriving at the equivalent action (\ref{properaction}) we have also provided the algorithm on how to obtain the coefficients, $c_{i,n}$'s, starting from a generic covariant action
that is regular as $\cR\ra 0$. Thus given any action of the form   (\ref{semi-final}) (and
indeed   (\ref{general})) we can determine an action of the form
  (\ref{properaction}) that is identical to the general action up to quadratic
order in fluctuations around dS/AdS/Minkowski background. It is worth emphasising that all the coefficients $c_i$'s depend on the background curvature
parameter $\Rb$, which is determined according to   (\ref{backgroundK}) from the original action.

Finally, we note that the Gauss Bonnet scalar being a topological invariant in
four dimensions allows us to set one of the coefficients among
$c_{1,0},c_{2,0},c_{3,0}$ to zero, if we wanted to.
This completes the derivation of the equivalent quadratic action in terms of
the curvature tensors. In the next section we will provide the perturbative
structure of action (\ref{properaction}) and the conditions for having a ghost and tachyon free spectrum around the dS/AdS space-times.
\section{Quadratic Fluctuations around dS/AdS/Minkowski background}
\subsection{Action \& Field Equations}
The goal of this subsection is to obtain the $\cO(h^2)$ action starting from the equivalent action   (\ref{properaction}) in a form that is suitable to address issues of stability and consistency. For this purpose, it becomes imperative that we not only find an expression for the $\cO(h^2)$ Lagrangian, but also that we decouple the Lagrangian into separate sectors containing the different physical degrees of freedom of the metric, and present it in a form where we can read off the corresponding propagators.  So, we will  have to decompose the metric tensor into its 10 degrees of freedom:
\be
h_{\mu\nu}=\hp_{\mu\nu}+\n_{(\mu}\Ap_{\nu)}+(\n_{\mu}\n_{\nu}-\4
g_{\mu\nu}\Box)B+\4 g_{\mu\nu}h \ ,
\label{decomposition}
\ee
where $\hp_{\mu\nu}$ represents the transverse traceless massless spin-two
graviton,
\be
\n^{\mu}\hp_{\mu\nu}=g^{\mu\nu}\hp_{\mu\nu}=0\ ,
\ee
containing 5 degrees of freedom, $\Ap_{\mu}$ is the transverse vector,
\be
\n^{\mu}\Ap_{\mu}=0\ ,
\ee
accounting for three degrees of freedom, and the two scalars $B$ and $h$
make up the remaining  two degrees of freedom. We should mention that in all the calculations that follow we will be using the $-+++$ signature for the metric.

Our next step is to substitute   (\ref{decomposition}) into   (\ref{properaction}) and simplify the Lagrangian to the point where we obtain decoupled actions for the different modes. It turns out that such a simplifying and decoupling process provides an extra-ordinary algebraic and technical challenge the details of which we provide in our companion paper~\cite{companion}. Here we present the main physical arguments and results. As noted earlier, a-priori, the metric represents a massless spin-2, a massless spin-1, and two scalar fields; three gauge degrees reduce the spin two field to
the two spin-two helicity states, while another gauge freedom can be used to
eliminate the time like component of the vector to again leave us with the two spin-one
helicity states.

Now, it is expected, and has been explicitly verified around the Minkowski background~\cite{Biswas:2011ar,Biswas:2013kla}, that only the spin-2 graviton and one of the scalar fields should survive. Indeed, one finds that when one substitutes the decomposed metric   (\ref{decomposition}) into the action (\ref{properaction}), all the terms involving the vector field, $\Ap_{\mu}$, automatically drops out. Also, only one combination of the two scalar fields,
\be
\phi\equiv\Box B-h\ ,
\ee
survive.

After a tour-de-force calculation, we obtain a radically simplified action:
\be
S= S_0+S_2+\cO(h^3)\ ,
\ee
where
\begin{eqnarray}
S_2&\equiv&\2\int dx^4\sqrt{-\gb}~{\hw}^{\perp\mu\nu} \LF \Box-\frac \Rb6 \RF \left[1+{4\over M_p^2} c_{1,0}\Rb+{2\over M_p^2}
\left\{\LF \Box-\frac\Rb6 \RF{\Fc}_2(\Box)\right. \right.\nonumber \\
&&+ \left. \left. 2\left(\Box-\frac
\Rb3\right)\Fc_3\left(\Box+\frac{\Rb}3\right)\right\} \right] {\hw}^\perp_{\mu\nu}\,,
\end{eqnarray}
and
\ba
S_0&\equiv&-\2\int dx^4\sqrt{-\gb}\w{\phi}\LF \Box+{\Rb\over 3}\RF\left[1+{4\over M_p^2} c_{1,0}\Rb-{2\over M_p^2}\left\{6\LF\Box+{\Rb\over 3}\RF\Fc_1(\Box)\right.\Rd\nonumber \\
&&+\left.\left.
\frac1{2}\Box{\Fc}
_2\left(\Box+\frac2 { 3 }
\Rb\right)\right\}
\right]\w{\phi}\ .
\ea
Here, we have introduced canonical fields
\ba
\hw^\perp_{\mu\nu}=\2M_p\hp_{\mu\nu}\mand
\w{\phi}=\sqrt{3\over 32}M_p\phi\ .
\ea

It is now straight forward to obtain the field equations
\ba
 \LF \Box-\frac \Rb6 \RF \left[1+{4\over M_p^2} c_{1,0}\Rb+{2\over M_p^2}
\left\{\LF \Box-\frac\Rb6 \RF{\Fc}_2(\Box)+  2\left(\Box-\frac
\Rb3\right)\Fc_3\left(\Box+\frac{\Rb}3\right)\right\} \right] {\hw}^\perp_{\mu\nu}&=&\ka \tau_{\mu\nu}\non
-\LF \Box+{\Rb\over 3}\RF\left[1+{4\over M_p^2} c_{1,0}\Rb-{2\over M_p^2}\left\{6\LF\Box+{\Rb\over 3}\RF\Fc_1(\Box)+
\frac1{2}\Box{\Fc}
_2\left(\Box+\frac2 { 3 }
\Rb\right)\right\}
\right]\w{\phi}&=&\ka \tau\non
\ea
where $\tau_{\mu\nu},\ \tau$ represents the appropriate stress-energy sources for the gravitational fields.
We have performed several checks of the above result in~\cite{companion}.
\subsection{Consistency Conditions}
The condition for the theory not to have any ghost/tachyon-like states around the Minkowski space-time was obtained in~\cite{Biswas:2011ar,Biswas:2013kla} by looking at the propagators. Although, essentially the propagators are the inverses of the field equation operators, obtaining its precise form  is somewhat of a technical exercise on dS/AdS space-times, see for instance~\cite{dS,AdS} for a discussion. For us, all we need to care about is the number and nature of the zeroes in the field equation operators for the tensor and scalar modes respectively:
\ba
\cT(\Rb,\Box)&\equiv&  \LF \Box-\frac \Rb6 \RF \LT 1+{4\Rb\over M_p^2} c_{1,0}+{2\over M_p^2}
\left\{\LF \Box-\frac \Rb6
\RF{\Fc}_2(\Box)+2\left(\Box-\frac
\Rb3\right)\Fc_3\left(\Box+\frac{\Rb}3\right)
\right\}\RT \ ,\non
\cS(\Rb,\Box)&\equiv& -\LF \Box+{\Rb\over 3}\RF\LT 1+{4\Rb\over M_p^2}c_{1,0}-{2\over M_p^2}\left\{2(3\Box+\Rb)\Fc_1(\Box)+
\frac1{2}\Box{\Fc}
_2\left(\Box+\frac2 { 3 }\Rb\right)\right\}\RT\ .\non
\ea

To see this, let us first look at the GR operators:
\ba
\cT_{\mt{GR}}(\Rb,\Box)&\equiv&  \LF \Box-\frac \Rb6 \RF \ ,\non
\cS_{\mt{GR}}(\Rb,\Box)&\equiv& -\LF \Box+{\Rb\over 3}\RF\ .
\ea
As is evident, the function, $\cT_{\mt{GR}}$, has a zero at $\Box=\Rb/6$, corresponding to a pole in the propagator that is known to represent the massless graviton state, the ``artificial'' mass is simply an artifact of the non-zero curvature of dS/AdS. $\cS_{\mt{GR}}$ also possesses a zero at $\Box=-\Rb/3$, and a corresponding pole in the propagator. As in the Minkowski case, the ghost-like scalar state (note the negative sign in front of the $\cS_{\mt{GR}}$ operator) is again needed to cancel the unphysical longitudinal degrees of freedom in the graviton field.

Let us now focus on the general $\cT(\Rb,\Box),\ \cS(\Rb,\Box)$ functions. Firstly, we recognize the presence of the zeroes representing the graviton and scalar modes that are present in normal GR. Secondly, just as in the Minkowski case, to ensure that we do not introduce a Weyl ghost in the tensorial mode, we must impose that there are no extra zeroes in $\cT(\Rb,\Box)$, or equivalently,
\be
a(\Rb,\Box)\equiv   1+{4\Rb\over M_p^2} c_{1,0}+{2\over M_p^2}
\left[\LF \Box-\frac \Rb6
\RF{\Fc}_2(\Box)+2\left(\Box-\frac
\Rb3\right)\Fc_3\left(\Box+\frac{\Rb}3\right)
\right]
\ee
should not have any zeroes. Finally, again as in the Minkowski case, the scalar function, $\cS(\Rb,\Box)$ can have one extra zero, as that would correspond to a pole in the propagator which will have  the correct residue sign. Indeed this zero  corresponds to the Brans-Dicke scalar degree of freedom. Thus, the function,
\be
b(\Rb,\Box)\equiv  1+{4\Rb\over M_p^2}c_{1,0}-{2\over M_p^2}\left[2(3\Box+\Rb)\Fc_1(\Box)+
\frac1{2}\Box{\Fc}
_2\left(\Box+\frac2 { 3 }\Rb\right)\right]
 \ ,
\ee
can at least have a single zero. If $b(\Rb,\Box)$ does contain a zero, then one has to ensure that the resulting scalar degree of freedom is not tachyonic:
\be
\mx{If }b(\Rb,m^2)=0\mx{ then }m^2>-{\Rb\over 3}\ .
\ee

Several comments are now in order:
\bi
\item The conditions that we obtained obviously reduces to the conditions that were previously enumerated for the Minkowski case in~\cite{Biswas:2011ar,Biswas:2013kla} when $\La\ra 0$.
\item It is appropriate to point out a particular special case where $\cF_2=\cF_3=0$ and $\cF_1=c_{1,0}$, a constant. In this case the tensor mode does not get any correction from it's GR counterpart, but the scalar propagator picks up an extra pole. This is indeed the Brans-Dicke scalar mode that appears in the Starobinsky inflationary model~\cite{Starobinsky}.
\item It should be apparent that both the scalar and tensor propagators depend on the background curvature, and thus if a particular model of gravity admits more than one constant curvature background, it is possible that the theory is consistent on one background and not the other. To view it differently, requiring that a theory of gravity be consistent around all possible backgrounds may be a powerful way to narrow down the list of acceptable theories of gravity.
\ei
\section{Discussion}

To summarise, here we have provided a formalism on how to find a quadratic (in curvatures) order action
of gravity that is equivalent to any given covariant gravitational action as far as linearized fluctuations are concerned around constant curvature maximally symmetric space-times. As elaborated in~\cite{companion}, while perturbing the quadratic curvature action  around dS/AdS/Minkowski metrics we found that only the spin-2  massless graviton and possibly a spin-0 Brans-Dicke scalar can propagate in these backgrounds.
We also enumerated the conditions under which the theory can be made perturbatively stable, i.e. the conditions for a given theory to be free from ghosts and tachyons. Our results match the limits of  Minkowski space-time~\cite{Starobinsky} for  quadratic curvature gravity with infinite derivatives, as well as the limit of pure Einstein-Hilbert action on dS/AdS backgrounds.

While our analysis can be applied to obtain viable cosmological models involving inflationary or bouncing cosmology as well as the modified gravity models motivated by the cosmic speed-up problem, it also provides encouraging signs for efforts in constructing a more fundamental gravity model which is bereft of the UV problems of GR. Classically, for the IDG theories, the next big step would be to be able to compute perturbations around cosmological and spherically symmetric solutions, because that would help us analyse a wide array of phenomenological applications that have made GR such a success. On the quantum front, while toy models have provided us with some encouraging results regarding finiteness of higher loops in infinite derivative theories~\cite{Talaganis:2014ida}, see also~\cite{Biswas:2004qu,BCK,BKR}, whether there is any chance that the higher loops can be made finite in IDG theories remains an intriguing question for future!

\begin{acknowledgement}
We would like to thank Spyridon Talaganis for discussions. TB would like to thank Carl for his insightful comments on the general subject matter of IDG theories. AM is supported by the STFC grant ST/J000418/1. AK is supported by the FCT Portugal fellowship SFRH/BPD/105212/2014 and in part by RFBR grant 14-01-00707.
\end{acknowledgement}

\end{document}